
%
%
\newbox\partialpage
\def\begindoublecolumns{\begingroup
 \output={\global\setbox\partialpage=\vbox{\unvbox255\bigskip}}\eject
 \output={\doublecolumnout} \hsize=14pc \vsize=89pc}

\def\doublecolumnout{\splittopskip=\topskip\splitmaxdepth=\maxdepth
 \dimen1=144pc \advance\dimen1 by-\ht\partialpage
 \setbox0=\vsplit255 to\dimen1 \setbox2=\vsplit255 to\dimen1
 \shipout\pagesofar \unvbox255 \penalty\outputpenalty}
\def\pagesofar{\unvbox\partialpage
 \wd0=\hsize \wd2=\hsize \hbox to\pagewidth{\box0\hfil\box2}}
\def\balancecolumns{\setbox0=\vbox{\unvbox255} \dimen1=\ht0
 \advance\dimen1 by\topskip \advance\dimen1 by-\baselineskip
 \divide\dimen1 by2 \splittopskip=\topskip
 {\vbadness=10000 \loop \global\setbox3=\copy0
  \global\setbox1=\vsplit3 to\dimen1
  \ifdim\ht3>\dimen1 \global\advance\dimen1 by1pt \repeat}
 \setbox0=\vbox to\dimen1{\unvbox} \setbox2=\vbox to\dimen1{\unvbox3}
\pagesofar}

\def\GeV{{\, \rm GeV\, }}
\def\MeV{{\, \rm MeV\, }}

\def\cm{{\, \rm cm\, }}

\def\boxit#1{\vbox{\hrule{\vrule\kern3pt
  \vbox{\kern3pt#1\kern3pt}kern3pt\vrule}\hrule}}

\newbox\bigstrutbox
\setbox\bigstrutbox=\hbox{\vrule height12pt depth5pt width0pt}
\def\bigstrut{\relax\ifmmode\copy\bigstrutbox\else\unhcopy\bigstrutbox\fi}

\def\la{\mathrel{\mathpalette\fun <}}
\def\ga{\mathrel{\mathpalette\fun >}}
\def\fun#1#2{\lower3.6pt\vbox{\baselineskip0pt\lineskip.9pt
  \ialign{$\mathsurround=0pt#1\hfil##\hfil$\crcr#2\crcr\sim\crcr}}}

\font\tenrm=cmr10 scaled 1200
\font\sevenrm=cmr7 scaled 1200
\font\fiverm=cmr5 scaled 1200
\font\teni=cmmi10 scaled 1200
\font\seveni=cmmi7 scaled 1200
\font\fivei=cmmi5 scaled 1200
\font\tensy=cmsy10 scaled 1200
\font\sevensy=cmsy7 scaled 1200
\font\fivesy=cmsy5 scaled 1200
\font\tenex=cmex10 scaled 1200
\font\tenbf=cmbx10 scaled 1200
\font\sevenbf=cmbx7 scaled 1200
\font\fivebf=cmbx5 scaled 1200
\font\tenit=cmti10 scaled 1200
\font\tensl=cmsl10 scaled 1200
\font\tentt=cmtt10 scaled 1200

\skewchar\teni='177 \skewchar\seveni='177 \skewchar\fivei='177
\skewchar\tensy='60 \skewchar\sevensy='60 \skewchar\fivesy='60

\font\ninei=cmmi10
\font\sixi=cmmi7
\font\fouri=cmmi5
\font\ninesy=cmsy10
\font\sixsy=cmsy7
\font\foursy=cmsy5

\skewchar\ninei='177 \skewchar\sixi='177 \skewchar\fouri='177
\skewchar\ninesy='60 \skewchar\sixsy='60 \skewchar\foursy='60

\def\twelvepoint{\def\rm{\fam0 \tenrm}
  \textfont0=\tenrm \scriptfont0=\sevenrm \scriptscriptfont0=\fiverm
  \rm
  \textfont1=\teni \scriptfont1=\seveni \scriptscriptfont1=\fivei
  \def\mit{\fam1 } \def\oldstyle{\fam1 \teni}
  \textfont2=\tensy \scriptfont2=\sevensy \scriptscriptfont2=\fivesy
  \def\cal{\fam2 }
  \textfont3=\tenex \scriptfont3=\tenex \scriptscriptfont3=\tenex
  \textfont\itfam=\tenit \def\it{\fam\itfam\tenit}
  \textfont\slfam=\tensl \def\sl{\fam\slfam\tensl}
  \textfont\bffam=\tenbf \scriptfont\bffam=\sevenbf
     \scriptscriptfont\bffam=\fivebf \def\bf{\fam\bffam\tenbf}
  \textfont\ttfam=\tentt \def\tt{\fam\ttfam\tentt}
  \baselineskip=14pt}

\outer\def\beginsection#1\par{\bigskip\vbox{\message{#1}\noindent{\bf#1}}
  \nobreak\smallskip\vskip-\parskip\noindent}
\def\exdent#1\par{\noindent\hang\frenchspacing#1\par}
\hsize=6.5in
\vsize=9in
\parskip=0pt

\def\m@th{\mathsurround=0pt}
\def\complex#1{\left\{\,\vcenter{\baselineskip=18pt\m@th
   \ialign{$\displaystyle{##}\hfil$&\quad$\displaystyle{##}\hfil$\crcr#1
   \crcr}}\right.}

\twelvepoint
\baselineskip=18pt
\def\erf{{\rm erf}}
\def\chiN{${\tilde \chi}N\;$}
\def\ge73{$^{73}$Ge}
\def\si29{$^{29}$Si}
\def\neut{${\tilde \chi}\;$}
\def\bino{${\tilde B\;}$}
\overfullrule=0pt

\line{\hfil {\bf UCRL-JC-114085}}
\line{\hfil {\bf UCSD/ASTRO-TH 93-02 }}
\bigskip
\bigskip
\centerline{\bf Nuclear Shell Model Calculations of}
\centerline{\bf Neutralino-Nucleus Cross Sections for \si29 and \ge73}
\bigskip
\bigskip
\centerline{\it M. Ted Ressell,$^{1,2}$ Maurice B. Aufderheide,$^{1,2}$
Stewart D. Bloom,$^1$}
\centerline{\it Kim Griest,$^3$ Grant J. Mathews,$^{1,2}$ David A. Resler$^4$}
\bigskip
\bigskip
\centerline{$^1$P-Division/Physical Sciences Directorate}
\centerline{Lawrence Livermore National Laboratory}
\centerline{Livermore, CA 94550}
\medskip
\centerline{$^2$Institute of Geophysics and Planetary Physics}
\centerline{Lawrence Livermore National Laboratory}
\centerline{Livermore, CA 94550}
\medskip
\centerline{$^3$Physics Department}
\centerline{University of California, San Diego}
\centerline{La Jolla, CA 92093}
\medskip
\centerline{$^4$N-Division/Physical Sciences Directorate}
\centerline{Lawrence Livermore National Laboratory}
\centerline{Livermore, CA 94550}
\vskip 0.35in
\baselineskip=14pt
\noindent{\bf Abstract.}
We present the results of detailed nuclear shell model calculations
of the spin-dependent elastic cross section for neutralinos
scattering from \si29 and \ge73.  The calculations were
performed in large model spaces which adequately describe the
configuration mixing in these two nuclei.  As tests of the
computed nuclear wave functions, we have calculated several nuclear
observables and compared them with the measured values and found
good agreement.  In the limit of zero momentum transfer, we find
scattering matrix elements in agreement with previous estimates
for \si29 but significantly different than previous work
for \ge73.  A modest quenching, in accord with shell model
studies of other heavy nuclei, has been included to bring agreement
between the measured and calculated values of the magnetic
moment for \ge73.  Even with this quenching, the calculated
scattering rate is roughly a factor of 2 higher than the best previous
estimates; without quenching, the rate is a factor of 4
higher.  This implies a higher sensitivity for germanium
dark matter detectors.  We also investigate the role of finite momentum
transfer upon the scattering response for both nuclei and find
that this can significantly change the expected rates.  We close
with a brief discussion of the effects of some of the
non-nuclear uncertainties upon the matrix elements.

\baselineskip=18pt

\vfill\eject

\noindent{\bf I. Introduction}
\bigskip

A host of astronomical evidence points to the existence of large
amounts of dark matter in the Universe [1].  Despite the
overwhelming amount of evidence for this dark matter's existence,
its exact nature remains a mystery.  Numerous candidates
have been proposed.  These include both
ordinary baryonic and non-baryonic matter [2].
Among the best motivated, and hence highly favored, of
non-baryonic candidates
is the lightest supersymmetric particle (LSP).  Experimental and theoretical
considerations suggest that the LSP is a neutralino, ${\tilde \chi}$,
made up of a linear combination of the photino, Z-ino, and 2 higgsinos (or
equivalently, the B-ino, neutral W-ino, and 2 higgsinos)
$${\tilde \chi} = Z_1 {\tilde B} + Z_2 {\tilde W_3} + Z_3
{\tilde H_1} + Z_4 {\tilde H_2}. \eqno(1) $$
The neutralino is an ideal dark matter candidate.  The
motivation for its existence arises naturally in modern theories of
particle physics [3], not as an {\it ad hoc} solution to the dark
matter problem.  For a very large region of
supersymmetric parameter space, neutralinos provide densities sufficient
to account for the mass-energy density of the Universe
[4].  The \neut also possesses
the virtue of potential detectability.  Its detection may be possible
in at least two ways: indirectly, through the products of
$\overline{\tilde\chi}
\tilde\chi$ annihilation from capture in the Sun or Earth
[5], or directly, via elastic
neutralino-nucleus, \chiN, scattering in a detector [6,7].
In either case, the \chiN elastic scattering cross section is
an essential ingredient.
In this paper we discuss detailed nuclear structure calculations
relevant to \chiN  scattering for two important elements,
silicon and germanium.

Neutralino-nucleus scattering is governed by physics at
several energy scales.  The mass and the mixing of the \neut,
and hence its interactions with quarks, are fixed near
the electroweak scale.  The spin-dependent interactions of
the \neut with
protons and neutrons are determined by the distribution
of quark spin within the nucleon, which depends upon physics at the
QCD scale.  There are also spin-independent contributions which
depend upon the quark content of the nucleons.
At the modest momentum transfers available to
dark matter particles $({\cal O}(\MeV))$ the \neut interacts with the entire
nucleus, not individual nucleons within it.  Thus, nuclear
structure plays an important role in determining
the strength of the \chiN  cross section.  The uncertainties
inherent in the electroweak and QCD-scale physics can be
parameterized in different ways.
The electroweak parameterization entails choosing
the exact composition and mass  of the \neut which, in turn,
is determined by parameters in a Lagrangian of a Supersymmetric
model.  At the QCD level
there are currently two competing possibilities for the coupling
of protons and neutrons to the \neut.  The currently favored
parameterization depends upon the measured quark spin
content of the proton determined by the European Muon
Collaboration (EMC) [8].  In contrast to these values of the
spin content are those derived from the Naive Quark Model
(NQM).  Experiments are being carried out which will
hopefully clarify this issue.
In this paper, we will investigate the effects of both the EMC
and the NQM estimates, as well as the uncertainties in their
determination.

At the nuclear level, several
attempts have been made to improve the cross section calculations.
Initial investigations into spin-dependent \chiN  scattering made
use of the independent single particle shell model (ISPSM)  in which
the character of the nucleus is completely determined by a
single unpaired valence nucleon [6,9,10].
Subsequent studies [11-15] have revealed a number of inadequacies of the
ISPSM and improved upon its estimates of the scattering
matrix element.  Engel and Vogel [11] used the odd group model
(OGM) and mirror pair $\beta-$decays, in an extended
odd group model (EOGM), to show that the ISPSM
can be off by a large factor when nuclei are far from
closed shells (see also ref. [12]).
Pacheco and Strottman [13] reached the same conclusion
by performing detailed nuclear shell model calculations for
several light nuclei.  The odd group and shell model treatments
obtain good agreement for light nuclei but, as the atomic mass
increases, nuclear configuration mixing produces cancellations
not considered in the OGM. This is most clearly evident
in $^{35}$Cl (the largest nucleus considered in ref. [13]) where
the OGM and shell model estimates of the matrix elements differ
by roughly a factor of 3 (see Table 1).  For comparison,
in Table 1 we have also included our calculation of the spin
in $^{35}$Cl using the same interaction and model space as
discussed for \si29 in Section IIa.  It is reassuring to note
that the two shell model calculations are in close agreement
despite the use of different interactions.  This effect has also been
seen in $^{93}$Nb [14] revealing the need for full shell model
calculations of heavier nuclei.

The need for more detailed calculations of the spin
content of heavier nuclei
is exacerbated by the fact that at least two of the most
promising detector programs currently under way are based
upon
\ge73 and \si29 [15].  Both of these nuclei are likely to have
large amounts of configuration mixing, making OGM estimates
suspect.  Furthermore, both \ge73 and \si29 require very large
model spaces. This is why the calculations
were not carried out in ref. [13].  In fact, in a recent review
[7] the statement was made that, ``...until a
reliable calculation in \ge73 has been carried out, the
description of the nuclear physics of dark matter detection will
be incomplete."

With these last comments as motivation, we have used
the nuclear shell model code CRUNCHER [16] to perform large basis
calculations of \ge73 and \si29 with the aim of providing
accurate cross sections for determining event rates for
spin-dependent scattering in the Ge and Si detectors currently
under development.

All of the work discussed above has been in the zero momentum
transfer limit.  As the experimental lower limit to the \neut  mass
continues to increase, finite momentum transfer
needs to be considered.  This is especially true for heavier nuclei,
such as \ge73 [7].  Unfortunately, phenomenological
models, such as the OGM and the Interacting Boson Fermion Model
(IBFM) [17] which have been used in the $q = 0$
limit cannot be easily extended to finite momentum transfer.
Furthermore, the ISPSM has been shown to be a poor approximation
at finite $q$ [7,14,18].  Fortunately, incorporating
finite momentum transfer into a full nuclear shell model
calculation is straightforward.  The formalism is
presented in [7]. It is a simple extension of that used in
semi-leptonic weak and electromagnetic interactions with
nuclei [19].
In Section IV, we show that finite momentum transfer
is important for both \si29  and \ge73.

\bigskip
\noindent{\bf II. The Zero Momentum Transfer Limit}
\bigskip

Neutralinos in the halo of our Galaxy can be characterized by
a mean virial velocity of,
$v \simeq \langle v \rangle  \simeq 300 {\rm km/sec} =
10^{-3} c$.  The maximum characteristic momentum transfer in \chiN
scattering will be $q_{max} = 2 M_r v$  where $M_r$ is
the reduced mass of the \chiN  system.  If the product
$q_{max}R$ is small ($\ll 1$), where $R$ is the nuclear size,
the matrix element for spin-dependent \chiN scattering reduces to a very
simple form [7,11]
$$ {\cal M} = A \langle N\vert a_p {\bf S}_p + a_n {\bf S}_n
\vert N \rangle \cdot {\bf s}_{\tilde \chi} \eqno(2) $$
where
$$ {\bf S}_i = \sum_k {\bf s_i (k)}, \;\;\;\; i = p,n \eqno(3)$$
is the total nuclear spin operator, $k$ is a sum over all
nucleons, and $a_p,\, a_n$ are
\neut-nucleon coupling constants which depend
upon the quark spin-distribution within the nucleons and on the
composition of the \neut, see e.g. eqs. (12) and (13) for the specific
cases of the \bino and ${\tilde \gamma}$.  Much of the uncertainties arising
from electroweak and QCD scale physics are encompassed by
$a_p$ and $a_n$.  The normalization $A$ involves the coupling
constants, masses of the exchanged bosons and various LSP
mixing parameters that have no effect upon the nuclear matrix
element.  To maintain contact with the previous literature [4,6,10,11,13], we
note that eq. (2) has often been written as
$$ {\cal M} = A \lambda \langle N\vert {\bf J}
\vert N \rangle \cdot {\bf s}_{\tilde \chi} \eqno(4a) $$
with
$$ \lambda = {{\langle N\vert a_p {\bf S}_p + a_n {\bf S}_n
\vert N \rangle}\over{\langle N\vert {\bf J}
\vert N \rangle}} =
{{\langle N\vert ( a_p {\bf S}_p + a_n {\bf S}_n ) \cdot {\bf J}
\vert N \rangle}\over{ J(J+1)
}}. \eqno(4b)$$
Examples of the full \chiN cross section can be found in the appendix
and in refs. [4,7,10,11]. Calculations
of the matrix element in eq. (2) are straightforward, although
computationally intensive, in the nuclear shell model.

Our approach to modeling the nuclei \si29 and \ge73 is quite
straightforward in principle.  Using a reasonable two-body
interaction Hamiltonian
we calculate the nuclear wave functions in an appropriate
model space.  As checks on these wave functions we
compute the excited state energy level spectrum, the magnetic
moments, and the single particle spectroscopic factors for each nucleus  and
compare them with experimentally measured
quantities.  Once reasonable agreement between the calculations
and measurements is obtained the ground state wave functions are used
to calculate the \chiN nuclear matrix element,
$\langle N\vert a_p {\bf S}_p + a_n {\bf S}_n
\vert N \rangle$ of eq. (2).  (We present our results using
the convention that all angular momentum operators are evaluated in their
$z$-projection in the maximal $M_J$ state, e.g.
$\langle {\bf S} \rangle \equiv \langle N \vert {\bf S} \vert N \rangle
= \langle J,M_J = J \vert S_z \vert J,M_J = J \rangle$.)  These
wave functions are also used to compute the more complicated
finite momentum transfer matrix elements.  In practice these steps
may be quite time consuming.  In the following two subsections
we detail the steps entailed in calculating the nuclear matrix elements
for spin-dependent \chiN scattering with \si29 and \ge73.

\bigskip
\noindent{\bf IIa. \si29}
\bigskip

For the nucleus \si29, the calculation of the nuclear wave
functions and matrix elements is almost as straightforward as
presented above,  because a well defined and tested
interaction in a well defined model space exists [20].  This
greatly facilitates the calculation.  In fact, after the
completion of these calculations, it was brought to our
attention that the calculation of \si29's spin had already been
done, using the same interaction and model space, but in a different
context [21].  In spite of this, we will detail our calculation
for two reasons.  First, the calculation of the \si29 matrix
element will serve as a useful comparison for the accuracy
of the calculation of \ge73's matrix element.  The \si29
computation will help to establish just how reliably one
can use the shell model to determine the relevant observables.
Second, we will show in a subsequent section that finite momentum
transfers are important for Si and this calculation
does not exist in the literature.

For our calculation of the wave functions for \si29 we have used
the USD interaction of Wildenthal [20] in a full {\it sd} shell model
space.  This interaction has been meticulously developed
and tested over many years and is known to accurately
reproduce numerous nuclear observables for {\it sd} shell nuclei.
Our computations were performed using the Lanczos method m-scheme nuclear
shell model code CRUNCHER  [16] and its auxiliary codes.
The m-scheme basis for \si29 in this model space has a dimension
of 80115 Slater determinants.  The calculation of the
wave functions takes roughly
a day and a half of time on a dedicated SPARC station.

In Fig. 1 we present the calculated vs. the
measured energy spectrum [22] for the 10 lowest eigenstates
of \si29.
It is clear that good overall agreement is achieved.

A test of more relevance to the \chiN scattering cross
section is the comparison between the predicted and measured ground-state
magnetic moment for \si29.  This is because of the similarity
of the relevant operators.  The magnetic moment, $\mu$, is given
by:
$$\mu =  \langle N\vert g_n^s {\bf S}_n + g_n^l {\bf L}_n +
g_p^s {\bf S}_p + g_p^l {\bf L}_p\vert N \rangle \eqno(5)$$
where the ${\bf S_i}$ are the total spin operators (in their
$z$ projection) from eq. (3), and the ${\bf L_i}$ are the
analogous orbital angular momentum operators.  The
{\it free particle} $g$-factors are given by:
$g_n^s = -3.826, \, g_n^l = 0, \, g_p^s = 5.586$, and
$g_p^l = 1$ (in nuclear magnetons).  There is ample evidence in shell model
calculations for many nuclei that these $g$-factors are quenched.  In \si29
this quenching is not necessary and is, in this context, counterproductive.
We will address this issue again in a
later section (quenching seems to be necessary to get $\mu$
``right'' in \ge73).  Using our wave functions we find
$\mu_{calc} = -0.50$, in agreement with the previous calculation
[21].  This is to be compared with the measured value
of $\mu_{exp} = -0.555$, which is good agreement by shell model
standards.  We also note that the ISPSM gives a value of
$\mu_{ISPSM} = -1.91$, in strong disagreement with experiment.

As a final check on the accuracy of our wave functions, we have
calculated the spectroscopic factors for a number of one
nucleon transfer reactions and compared them to the experimentally
determined values.  Spectroscopic factors $S(J;jJ_0)$ are defined
in [23] as
$$S(J;jJ_0) = \vert \langle \psi(J,M)\vert \Psi(j,J_0;J,M)\rangle\vert^2
\eqno(6)$$
where $\Psi(j,J_0;J,M)$ is the state obtained by coupling
an odd nucleon with angular momentum $j$ to the core
with angular momentum $J_0$ to obtain total angular momentum
$J$. $\psi(J,M)$ is the true nuclear wave function.  As
an example, we compute $S$ for the reaction
$^{28}{\rm Si} + d \longrightarrow {^{29}{\rm Si}} + p$
going from the ground state of $^{28}$Si, $J = 0$, to the ground
state of \si29, $J = 1/2$, in the ISPSM.  Since the incoming
neutron has spin 1/2 the only amplitude which can contribute
(in the ISPSM) is that in which the neutron is in a $L = 0$ state.
Thus, we have $S(1/2;1/2,0) = 1$.  (Note, experimental data is
generally presented as $(2 J + 1) S(J; j, J_0)$ and we will adhere
to that convention.)  In a full shell model calculation
(and real nuclei), configuration mixing complicates this simple picture.

To calculate the $S(J; j,J_0)$ for \si29, we have computed the
ground state wave function for $^{28}$Si using the same Hamiltonian
and model space.  As detailed in ref. [16] we then
act on this wave function with a neutron creation operator
and take the scalar product of the result with the relevant \si29 wave
function.
The most important calculation is, of course, the ground state
to ground state transition since it most directly probes the
nuclear configurations which can interact with the \neut.
As an additional check, we have calculated the spectroscopic
factors to the first two excited states as well.  The results
of these calculation are presented in Table 2 along with an
unweighted average of the comparable experimental data from Endt
[24]. We also present the ISPSM results.  Table 2 reveals the good
agreement between the full shell model calculation and experiment
as well as the inadequacy of the ISPSM.  This provides confidence
in our wave functions and our calculations of the \chiN matrix
elements.

Once we have suitable wave functions, the nuclear matrix element
in eq. (2) ($\langle N\vert a_p {\bf S}_p + a_n {\bf S}_n
\vert N \rangle$) is easily calculated for any set of
$a_p$ and $a_n$.  Here we compute
the spins in eq. (2), and then multiply them by the
desired $a_i$ to find the total matrix element.
In Table 3 we present the results of this
exercise using our wave functions.  Our results agree with
those found in ref. [21].  In Table 3 we also present the
results found in the ISPSM, the OGM, and the extended OGM
approach.  It is immediately apparent that the ISPSM is a very
poor approximation to the actual nuclear configuration.
Configuration mixing causes the total angular momentum to be
shared between the nuclear spin and orbital angular momentum,
contrary to the assumptions of the ISPSM (the total angular
momentum resides in the spin of the $2S_{1/2}$ neutron).  It is also apparent
that the OGM treatments and the shell model calculations are
in fair agreement as to the properties of \si29.  The important
point here is that the protons contribute negligibly to the
relevant nuclear properties, the key assumption of the OGM.  We will find,
however, that this assumption is violated for \ge73.

\bigskip
\noindent{\bf IIb. \ge73}
\bigskip

The availability of a well developed and tested interaction in
the {\it sd} shell made the calculation of the \si29 ground state wave
function relatively simple.  For our calculation of the ground
state of \ge73 we did not have the luxury of a good (in the sense
that it has been phenomenologically determined
and extensively tested) Hamiltonian and
there is not a well defined model space.  Hence there is more
uncertainty in determining a good
nuclear state which accurately reproduces the measured observables.
There is some evidence that \ge73
may be deformed [7] meaning that a large model space
may be required.
The extent to which our wave functions can be trusted
will be revealed by how well they
reproduce observable phenomena.
Our approach has been to choose fairly
large model spaces which should include most of the relevant
excitations and apply a fairly simple but well motivated interaction
to the nucleons within this space.  We then examine the ordering and
spacing of the excited state energy levels.  The single particle
energies of the interaction are then modified and the procedure
is repeated.  Thus, by a process of iteration, we have been able
to reproduce the low lying energy levels in \ge73.  This procedure
has been previously used with some success in this mass region [25].
As before, once a suitable fit to the energy spectrum
is obtained, the magnetic moment and spectroscopic factors are
calculated as additional checks on the wave function.  When we
are convinced that we have reasonable wave functions, we then
calculate the spin distribution.  We now present this
procedure in some detail.

In our study of \ge73, we have chosen to use the PMMA interaction
[26].
This interaction is an analytic approximation to the
Kallio and Kolltveit [27] interaction, which is a reasonable
approximation to a full $G$-matrix calculation. This interaction
has proven to be both adequate and tractable in
shell model calculations.  As discussed above,
the single particle energies were varied in order to match the
energy spectrum of low lying excited states.  Using this
interaction we have investigated two different model spaces.  The
first, which we will refer to as the small space, has a m-scheme
basis dimension of 24731 Slater determinants.
The other, creatively named the large
space, allows many more excitations with an
m-scheme basis dimension of 117137 Slater determinants.
Despite the fairly large size of the
bases, rather severe truncations in the space have been enacted.
The small space is the smallest in which we could obtain
agreement with the spectrum.  The large basis dimension was determined by
computer time and memory constraints.  Each iteration (of which
there were many) consumed roughly 4 days of SPARC time and several
hundred Mbytes of memory.

Each of these model spaces is ``customized'' to \ge73
and requires some explanation.  In standard notation the small
space may be described by:
$((1f_{5/2},2p_{3/2},2p_{1/2})^{16}$ $(1g_{9/2})^1$ $(2d_{5/2},1g_{7/2})^0 +$
$(1f_{5/2},2p_{3/2},2p_{1/2})^{16}$ $(1g_{9/2})^0$ $(2d_{5/2},1g_{7/2})^1 +$
$(1f_{5/2},2p_{3/2},2p_{1/2})^{15}$ $(1g_{9/2})^2$ $(2d_{5/2},1g_{7/2})^0)$
In words: the 16 particles in the $1f_{5/2},2p_{3/2}$, and $2p_{1/2}$
shells may ``move'' freely about in these shells,  there is also
1 particle (a neutron) in the $1g_{9/2}$; additionally, excitations
of the $1g_{9/2}$ neutron into the $2d_{5/2}$ or $1g_{7/2}$ shells
is allowed, {\bf or} 1 particle from the $1f_{5/2},2p_{3/2}$, and $2p_{1/2}$
shells may be excited into the $1g_{9/2}$ orbital (there
it can pair with the already present neutron).  In the same notation, the large
model space allows configurations given by:
$((1f_{5/2},2p_{3/2})^{14}$ $(2p_{1/2},1g_{9/2})^3$ $(2d_{5/2},1g_{7/2})^0 +$
$(1f_{5/2},2p_{3/2})^{14}$ $(2p_{1/2},1g_{9/2})^2$ $(2d_{5/2},1g_{7/2})^1 +$
$(1f_{5/2},2p_{3/2})^{13}$ $(2p_{1/2},1g_{9/2})^4$ $(2d_{5/2},1g_{7/2})^0)$.
The major difference here is that it is now possible to have up to
4 particles in the $1g_{9/2}$ orbital.

In Fig. 2 we show the excited state
energy spectrum of the first 10 calculated states
compared to the measured spectrum in the same
energy range for both the large and the small spaces.
In the small space we were able to iterate until
essentially perfect agreement between theory and experiment
was obtained for the lowest lying  five excited states.  There is then
a large region where there are many observed states but
no calculated states.  This is because our limited
model space is not large enough to contain all of the 2-particle
1-hole excitations necessary to describe such states.  The
omission of these seniority 3 configurations, from the ground
state, is probably not too serious.
At excitation energies $\ga$
1 MeV there are again several calculated states which
seem to have analogs in the measured spectrum of states.

With the greater number of available excitations in the large
space one would expect to start describing states in the spectrum between
0.5 MeV and 1 MeV, and this is observed for the
10 lowest lying states, as
demonstrated in the right side of Fig. 2.
In the large space it was
impractical and unnecessary to iterate until `perfect' agreement
was obtained for the lowest energy states (compare the scales of
Figs. 1 and 2 and the relative errors).  We did iterate until the
small changes in the single particle energies needed to move
the states no longer affected the spin matrix elements significantly.
Figure 2 reveals reasonable agreement between our
low lying calculated states
and the observed distribution of states giving us reasonable
assurance that we have
obtained a good description of the \ge73 ground state.
We now apply our other tests in the hope of confirming this
statement.

As previously noted, the most relevant test for the spin matrix elements is the
comparison of the magnetic moments.  Here our calculations
do an acceptable job but some quenching of the $g$-factors
is required to obtain complete agreement with experiment.
All other attempts
to calculate the matrix elements have relied heavily upon
quenching as well: the IBFM [17] and the OGM
[7,11] (which can be viewed
as a well motivated prescription for quenching the ISPSM value
of $\mu$).  In fact, the quenching required by our wave functions
is significantly less than in either of the above methods.
We will discuss the issue of quenching in a subsequent section.
The measured
magnetic moment of \ge73 is $\mu_{exp} = -0.879$ which is significantly
different from the ISPSM value of $\mu_{ISPSM} = -1.91$.
The values we calculate (without quenching) are:
$\mu_{ss} = -1.468$ for the small space
and $\mu_{ls} = -1.239$ for the large space, a major
improvement over the ISPSM, especially for the large space.
The main reason for the decrease in $\mu$ can be seen in Table
3.  Due to configuration mixing, the protons significantly
contribute to the total angular momentum, especially through
their orbital part.  Referring to eq. (5) it is obvious that a
large value of $\langle {\bf L}_p \rangle$ can have a significant effect
on $\mu$.  This large value for ${\bf L}_p$ is one of the
major results of the present work and the primary reason that our
answers will differ from those of previous analyses.
While one might wish for
better agreement between the measured and calculated magnetic
moments, by typical shell model standards [28,29], reasonable
agreement has been achieved.  Thus we continue to have confidence
in our results.

The final test we have performed for \ge73 is to compare
the calculated and measured spectroscopic factors, eq. (6),
for the reaction $^{72}{\rm Ge} + d \longrightarrow {^{73}{\rm Ge}}
+ p$ [30].  As with the magnetic moment comparison, we find quite
reasonable results, albeit not without some level of ambiguity.
We present our results, the experimental data, and the ISPSM
values for the three lowest lying states in Table 4.
For the ground state to ground state reaction, labeled
$1g_{9/2}$ in Table 4,  good
agreement is obtained between the computed and experimental value
for both model spaces.  Since this is the reaction which
directly tests the state we are interested in, this is quite
encouraging.  The small space obviously does not allow enough
excitations to correctly describe the excited state transitions.
The large space does a much better job on the excited states
but if one were interested in their properties, a better
description might be desirable.  Fortunately, we are
only interested in the ground state of \ge73 which seem to be
adequately described in these model spaces.

After examining the comparisons of the experimental and
calculated values of the excited state spectra, magnetic
moments, and spectroscopic factors we believe that we have
obtained a reasonable description of the \ge73 ground state.
The case for this is not unassailable but our wave functions
do include pieces of physics which other analyses have neglected
and they do reproduce a number of observables with good accuracy.
Having established the viability of our calculation of the
ground state wave function of \ge73, we now discuss
\ge73--\neut scattering.

The primary results of this section have already been presented
in Table 3,  which shows our results for the spin and orbital
pieces of total angular momentum (in their $z$-projection).  We
compare our results to those found in the ISPSM,
the OGM, and the IBFM.  We will confine our comments
henceforth to the large space calculation since it seems
to be a better representation of the ground state.  The small
space serves primarily to illustrate the sensitivity to
configuration mixing. The fact that an increase in the size of the
model space leads to an improvement in the magnetic moment is
especially heartening.  Table 3 shows that these
effects are small but finite.  As a final comment regarding Table 3
before we discuss its contents, we point out that:
there are two rows for the IBFM, in the first row the required
quenching should be applied to $a_p$ and $a_n$ in the matrix elements
(this corresponds to quenching the $g$-factors for $\mu$),
in the second the same quenching has, instead,
 been applied to the spins [7] (here the free particle
$g$-factors would be used for the magnetic moment).
These two approaches are equivalent but help to highlight the
different results obtained in the IBFM and shell model approaches.

The most interesting feature of Table 3 is the large contribution of
the proton's orbital angular momentum found in our calculations.
This large value of $\langle {\bf L}_p \rangle$ results in significant
differences in the spin matrix elements found in this investigation
vs. those found in the OGM and IBFM approaches.  The proton's
orbital angular momentum makes a major contribution to the magnetic
moment which is completely neglected in the OGM and not adequately
represented in the IBFM.  Surprisingly (and coincidentally), our
results for the spin tend to agree with those found in the ISPSM.
A large contribution to the angular momentum by the protons
violates the assumptions of the OGM. This explains the
factor of 2 difference in $\langle {\bf S}_n \rangle$ between the OGM
and our approach. (In a shell model treatment of $^{93}$Nb, a
similar result was found, except there it is the neutrons carrying
a large amount of orbital angular momentum which, once again,
invalidates the OGM [14].)  The IBFM wave functions do not
possess a large proton contribution to the angular momentum
of the nucleus and hence produce an unquenched magnetic moment
of $\mu_{IBFM} = -1.785$, similar to the single
particle model.  This is to be contrasted with the shell model
value of $\mu = -1.239$.  Thus, in the IBFM a very large quenching
factor of 0.523 must be applied to the spin matrix elements
(i.e. either $a_i$ {\it or} $\langle {\bf S}_i \rangle$ as is shown
in Table 3) in
order to match the measured magnetic moment of $\mu_{exp} = -0.879$.
If the same procedure is applied to the shell model matrix
elements (a procedure we do not recommend, see the next section)
a quenching factor of only 0.792 is required.  This is considerably less
quenching of the spin than in the IBFM.  If we apply this quenching
factor to the large space spins in Table 3 (in the same manner as was
done to create the second IBFM row), we find $\langle {\bf S}_n \rangle
= 0.371$ and $\langle {\bf S}_p \rangle = 0.009$, still quite
different from the OGM and the IBFM (and now the ISPSM) results.
Thus, we find that a full shell model calculation using realistic
wave functions obtains significantly different values for the spin
matrix elements than previous analyses.  In fact, our results
could result in up to a {\bf four-fold} increase in the \chiN scattering
rate over the estimates in previous work. This implies a much
greater sensitivity for germanium dark matter detectors.
We are now ready to
investigate the scattering response at finite momentum transfer
but first we address the issue of quenching in more detail.

\bigskip
\noindent{\bf III. Quenching of the Spin Matrix Elements}
\bigskip

The concept of quenching of the magnetic moment $g$-factors and
Gamow-Teller strength function in shell model calculations has a long
history.  What is clear is that quenching is almost
universally required once nuclei are filled beyond the
p-shell.  To quote from a thorough investigation of the M1
(magnetic moment) operator in the {\it sd} model space [28], ``There are
several reasons why even `perfect' shell-model calculations
of electromagnetic matrix elements based on the free-nucleon
characteristics of the neutron and proton should differ from the
corresponding experimental values.  In reality, the nuclear wave
functions must be more complicated than those of the theoretical
model we use.  Real nuclear states must involve nucleonic degrees
of freedom beyond the {\it sd}-shell space.  In addition, non-nucleonic
degrees of freedom which involve delta isobars and mesons may
be important in the observed phenomena.''
This `excluded' physics also plays a role in weak interaction
calculations [31].  The preferred method for dealing
with these inadequacies of the shell model is to quench the relevant
$g$-factors and so construct an effective magnetic moment
operator.  In the {\it sd} [28] and {\it fp} [29]
shells, systematic studies have been made to determine the
values of modified $g$-factors which give the best fit
to the magnetic moment for a large number of nuclei in each
shell.  In the {\it sd} shell study, the operator
$({\bf {\cal Y}}^{(2)} \otimes {\bf s})^{(1)}$ was shown to make a significant
contribution to the effective magnetic moment operator making comparisons
between $\mu$ and the \chiN matrix element somewhat more
ambiguous.
We can use these optimized $g$-factors to estimate the amount
of quenching we expect for our calculations of \chiN scattering on
\si29 and \ge73.

As mentioned above, quenching in the {\it sd}-shell with the
USD interaction has been thoroughly investigated and an
optimum set of quenched $g$-factors has been found
[28].  Following Table 3 of [28]
the optimized $g$-factors for \si29 should be:
$g_n^s = -3.24, g_p^s = 4.75, g_n^l = -0.091$, and $g_p^l = 1.129$.
We note that these spin $g$-values are consistent with an overall
spin quenching factor of $\sim 0.85$.  However, we see that it is not
only the spin $g$-factors which are modified.  Using these
values of the $g$-factors in eq. (5) leads to the following
magnetic moment $\mu(quenched) = -0.45$, slightly worse than
obtained with no quenching.  This is not surprising, since the
above values of the $g$'s were obtained as a global fit to
many {\it sd}-shell nuclei besides \si29.  Since the wave function
for \si29 already
matchs the measured magnetic moment it
is not surprising that quenching damages the agreement.  In
view of this we do not advocate any quenching for the
\si29 matrix elements.

As with all other aspects of this investigation, things become
considerably more complicated when we consider \ge73.
Due to the complexity of the nuclei in this mass region, no systematic
studies of quenching have been done.  The nearest relevant
study of which we are aware involves nuclei in the lower {\it fp} shell with
$41 \le A \le 49$ [29].  In this work, three different
interactions are compared and three very different results
emerge for the quenched $g$-factors.  Additionally, all three
estimates of the optimized $g$-factors result in improved agreement
between the measured and calculated magnetic moment of \ge73.  For
purposes of comparison, in Table 5 we show the free particle
$g$-factors, the 3 sets of quenched $g$-factors from ref.
[29], and the resultant magnetic moment of \ge73 obtained using
the large space wave function.  The key point of Table 5 is that
changes in one of the $g$'s can be compensated for by changes in
another and the choice of the best set is dependent upon the interaction
chosen.  Table 5 reveals that the effective
$g$-factors from the KB1 [29]
interaction produce the best fit to $\mu_{exp}$ for \ge73.
These $g$-factors result from an almost exclusive quenching of the
{\it isovector} part of the spin (isovector quenching factor
= 0.86).

There is one additional piece of information which may be
used to investigate the required quenching,
namely Gamow-Teller (GT) $\beta$-decays.  As with the magnetic moment
operator, the GT operator ($\propto g_A \tau_{\pm} {\bf S}$)
requires quenching when comparing shell model GT strengths
to the measured values [31].  Since the
GT operator is directly related to the isovector part of the spin,
it is a further probe of the quenching of the
isovector spin.  Several studies have shown that the free
nucleon value of $g_A = 1.25$ is {\it on average} reduced
to a value of $g_A \simeq 1$ in heavy nuclei.  This corresponds
to an isovector spin quenching factor of 0.8.  This is in
reasonable accord with the isovector quenching of the KB1
interaction discussed above and the similar quenching factor
found in the {\it sd} shell near $A = 28$, ($4.00/4.76 = 0.84$)
[28].  Finally, we note that this GT quenching
plays a role in the mirror pair analysis of Engel and Vogel
(through their ratio $R$) [11].  The issue of quenching is
thus quite general and quite troublesome when dealing with spin
matrix elements of heavy nuclei.

The calculated magnetic moment for
\ge73 with free particle $g$-factors already shows significant
improvement in agreement with measured values over
previous treatments due simply to the inclusion of configuration
mixing.  If {\it exact} agreement is desired, an effective operator,
with quenched $g$-factors must be invoked.
If quenching is invoked, we advocate quenching
only the isovector spin part.  Following this prescription
a quenching of the isovector spin $g$-factor by a factor
of 0.833 produces a calculated magnetic moment (large space)
of $\mu = -0.879$, in perfect agreement with experiment.  This
quenching factor is in agreement with the expectations of GT
decay in heavy nuclei.  In addition, it is consistent with the
isovector quenching found in the {\it sd} [28] and {\it fp} [29]
shells when fitting to magnetic moment data.  By following
this prescription we follow the path with greatest motivation and
obtain agreement with experiment by quenching only one
number, the isovector component of the spin.  Using
this quenching factor we find effective spin $g$-factors
of $g_n^s(eff) = -3.04 = 0.795 g_n^s(free)$ and
$g_p^s(eff) = 4.80 = 0.859 g_p^s(free)$.
In the zero momentum transfer limit
this quenching is equivalent to the large space matrix
elements becoming $\langle {\bf S}_n \rangle = 0.372$ and
$\langle {\bf S}_p \rangle = 0.009$ (when combined with the free
particle $g$-factors). These values should be
viewed as lower limits to the spin-matrix elements.
These values remain quite
different from any of the previous estimates performed in
the ISPSM, OGM, or IBFM.

In this section we have shown that a simple quenching
of the isovector-spin matrix element by a factor of
0.833 produces agreement with the measured magnetic
moment data for \ge73.  A quenching factor of this type,
and roughly this magnitude, could have been predicted based upon
previous shell model studies [28,29,31].  Given the amount
of evidence which supports this quenching, we feel that the
quenched values for the spin matrix elements are the ones
most likely to be correct.  Hence in calculating the \ge73--$N$
scattering cross section, one should use the values in the last row
of Table 3 (labeled ``large,quenched") and the normal $a_n$ and
$a_p$ {\bf or} use the values in the row above that with a quenching
factor of 0.833 applied to the isovector combination of $a_n$ and
$a_p$ while the isoscalar combination remains unchanged.
[To be more explicit: Calculate $a_1 = a_p - a_n$ and
$a_0 = a_p + a_n$ where nothing has yet been quenched.
Using these values of $a_0$ and $a_1$ recalculate $a_p$
and $a_n$ but now quench the isovector piece,
$a_p = {1\over{2}}(a_0 + 0.833 a_1)$ and
$a_n = {1\over{2}}(a_0 - 0.833 a_1)$.  These quenched
values of the $a_i$'s may now be combined with the values of the
spin in second to last line of Table 3 to calculate the total
matrix element.]
Having addressed the issue of
quenching, we may now consider
\neut scattering with \si29 and \ge73 at finite momentum
transfer.  For simplicity,
at finite momentum transfer, any quenching will
be applied to $a_n$ and $a_p$, as described in the square
brackets above or in the appendix, and not to the spin matrix elements.

\bigskip
\noindent{\bf IV. Finite Momentum Transfers}
\bigskip

When the LSP was first proposed as a viable dark matter
candidate, its preferred mass was between 5 and 10 GeV [32].
With a mass of this order and a typical galactic halo
velocity ($v \simeq 10^{-3} c$), the neutralino's total
momentum ($q \sim M_r v \sim 10 \MeV$) was small compared
to the inverse of the nuclear size ($1/R \sim 1/1{\rm fm} \sim
200 \MeV$) and the zero momentum transfer limit was appropriate
for studies of \chiN scattering.  Since then,
experiments at accelerators have pushed
the allowed \neut mass, $m_{\tilde \chi}$, to larger values (there
are ways around this if some of the theoretical assumptions are
relaxed [33]),  and it has been shown that
heavy \neut's are just as viable as a dark matter candidate as
the lighter ones[4,34].  As $m_{\tilde \chi}$
becomes larger than a few 10's of GeV the product
$qR$ starts to become non-negligible and finite momentum transfer
{\bf must} be considered for heavier nuclei.  (The maximum allowed
momentum transfer is $q_{max} = 2 M_r v$.)  In ref. [7]
a simple set of rules is given for when finite momentum
transfers are {\bf not} important:  i.e. the product $q_{max}R < 1$ if
(a) $A < 28$ or (b) $m_{\tilde \chi} < (1/(1.2 A^{4/3} + 100))\GeV$.
Thus, we see that finite momentum transfers may be important
in \si29 and are quite likely to be important for \ge73.
In this section we will show that these effects are
important for both \si29 and \ge73.  For spin-independent
scattering these effects have been considered in refs. [9,4].  For
spin-dependent scattering a simple approximation for
finite momentum transfer was applied in [12].

The formalism for elastic \chiN scattering at all momentum
transfers has been developed in refs. [7,18].  It is a
straightforward extension of the formalism developed for the
study of weak and electromagnetic semi-leptonic interactions
in nuclei [19].  The differential
\chiN cross section is given by:
$${{d \sigma}\over{d q^2}} = {{8 G_F^2}\over{(2J + 1) v^2}} S(q),\eqno(7)$$
where $S(q)$ is the spin structure function
$$S(q) = \sum_{L\, odd} \big( \vert\langle N \vert\vert {\cal T}^{el5}_L
(q) \vert\vert N \rangle\vert^2 +
\vert\langle N \vert\vert {\cal L}^5_L
(q) \vert\vert N \rangle\vert^2\big) . \eqno(8)$$
${\cal T}^{el5}(q)$ and ${\cal L}^5(q)$ are the transverse
electric and longitudinal multipole projections of the
axial vector current operator [19].
For their explicit form in the \neut context, see [7,18].
The reduced matrix elements of the multipoles in eq. (8) are
easily evaluated in the harmonic oscillator basis in the
nuclear shell model [19].
In the limit of zero momentum transfer $S(q)$ reduces to
[7,18]
$$S(0) = {1\over{4 \pi}} \vert\langle N \vert\vert\sum_i
{1\over{2}}(a_0 + a_1 \tau_3^i) {\bf \sigma}_i\vert\vert N
\rangle\vert^2\eqno(9a)$$
or, equivalently
$$S(0) = {1\over{4 \pi}} \vert (a_0 + a_1)\langle N \vert\vert
{\bf S}_p\vert\vert N\rangle + (a_0 - a_1)\langle N \vert\vert
{\bf S}_n\vert\vert N\rangle\vert^2\eqno(9b)$$
where $a_0$ ($a_1$) is the isoscalar (isovector) projection
of the $a_n$ and $a_p$ of eq. (2) ($a_0 = a_n + a_p$
and $a_1 = a_p - a_n$).  To make the connection with the zero
momentum transfer case even clearer, we can ``unreduce''
the matrix elements in eq. (9b) and rewrite it as
$$S(0) = {1\over{\pi}} f(N) \vert a_p \langle N \vert
{\bf S}_p\vert N\rangle + a_n \langle N \vert{\bf S}_n\vert
N\rangle\vert^2\eqno(9c)$$
where the matrix elements in eq. (9c) now correspond to the
entries in Table 3.  The factor $f(N)$ accounts for the
Wigner 3-$J$ function which relates the reduced and
unreduced values of the matrix elements and has the values
$f$(\si29) = 6 and $f$(\ge73) = 12.222.
Thus, eq. (8) reduces to the correct
expression in this limit.  As an aid in discussing the most
general neutralino state, $S(q)$ may be split into a pure
isoscalar piece, $S_{00}$, a pure isovector piece,
$S_{11}$, and an interference term, $S_{01}$, in the following
way:
$$S(q) = a_0^2 S_{00}(q) + a_1^2 S_{11}(q) + a_0 a_1 S_{01}(q). \eqno(10)$$

Two factors contribute to the maximum
allowed momentum transfer.  As $m_{\tilde \chi}$ becomes much greater
than the nuclear mass, $m_N$, the reduced mass asymptotes
to $M_r \rightarrow m_N$.  Also, the \neut's have a
Maxwellian velocity distribution in the halo and some will
possess velocities significantly greater
than $\langle v \rangle \simeq 10^{-3}c$.
A maximum velocity
of $v_{max} \simeq 2 \langle v \rangle$  implies maximum
momentum transfers of $q_{max}$(\si29) $\simeq 108\MeV$ and
$q_{max}$(\ge73) $\simeq 271\MeV$.  Neither of these values is
{\it small} compared to the inverse nuclear size.  In a harmonic
oscillator basis, the fiducial nuclear size is set by the
oscillator parameter,
$b = 1\,{\rm fm}\, A^{1/6} = (1/197.327 \MeV) A^{1/6}$
[23].  Hence we have values
$b$(\si29)$ \simeq 1.75\,$fm and $b$(\ge73)$ \simeq 2.04\,$fm.
These values result in the products: $(q_{max}b)_{\rm Si} = 0.96$
and $(q_{max}b)_{\rm Ge} = 2.80$. The effects of non-zero momentum
transfer might play an important role in these nuclei (depending
upon the \neut mass).  The nuclear multipole matrix elements in
eq. (8) are most easily evaluated and discussed in terms of the
quantity $y \equiv (bq/2)^2$.  We will use both $y$ and $q$
in our discussion.  The corresponding maximum values of $y$ are:
$y_{max}$(\si29)$ = 0.23$ and $y_{max}$(\ge73)$ = 1.96$.

Having determined the relevant range of momentum transfers to
investigate, we may now discuss the results obtained for the
spin structure function, $S(q)$ of eqs. (8)-(10), using our
wave functions for \si29 and the large space wave functions for
\ge73.  In Figs. 3 and 4 we display the functions
$S_{00}(y), \, S_{11}(y)$, and $S_{01}(y)$ for \si29 and
\ge73 respectively.  With these functions it is a simple
matter to calculate $S(q)$ for any \neut once the relevant
$a_0$ and $a_1$ are known.  As an aid in this endevour, we have made
simple exponential fits to $S(q)$ for both \si29 and \ge73.
We find that for \si29
$$S^{fit}_{Si}(y) = 0.00818 \big( a_0^2 e^{-4.428 y} + 1.06 a_1^2
e^{-6.264 y} - 2.06 a_0 a_1 e^{-5.413 y} \big) \eqno(11a)$$
provides a highly accurate fit to $S(y)$ for $y < 0.15$.  Beyond
this value of $y$ eq. (11a) begins to seriously underestimate the
value of $S(y)$.  The equivalent function for \ge73 is
$$S^{fit}_{Ge}(y) = 0.20313 \big( 1.102 a_0^2 e^{-7.468 y} + a_1^2
e^{-8.856 y} - 2.099 a_0 a_1 e^{-8.191 y} \big), \eqno(11b)$$
valid for $y < 0.2$.  These two fits should give reasonable
values for $S(q)$ in their region of validity for most
neutralinos.  We note that, if we assume $q = M_r v$,
$y = 0.1$ corresponds to $m_{\tilde \chi} \simeq 600 \GeV$
for \ge73, already at the upper end of interesting \neut masses.
Thus, eqs. (11) lack of validity at high $y$ $(q)$ should
not be a serious impediment to their use.  A more serious
problem with the fits in eq. (11) occurs when $a_p \gg a_n$
(or equivalently, when $a_0 \simeq a_1$).  In this regime of
\neut parameter space, there is a near cancellation of
$S_{00}$ and $S_{11}$ against $S_{01}$ and the fits do not
accurately reproduce the full result.  This is a problem for the
NQM values of $a_n$ and $a_p$ for a pure \bino, for example
(these are given below).
For a true representation of $S(q)$ in this case, one must
use the full shell model results as presented in Figs.
3 and 4 for all $y$.  With that warning, the fits in eq. (11) should
be quite useful for determining $S(q)$ for the great majority
of \neut cases.  Quenching is implemented in eqs. (11) by making
the substitution $a_1 \longrightarrow 0.833 a_1$.

In Figs. 5 and 6 we display $S(q)$ for a pure \bino ($ Z_1 = 1,
Z_2 = Z_3 = Z_4 = 0$).  Theoretical arguments lead one to
believe that an almost pure \bino (purity $>$ 99\%) is the most
favored \neut composition once $m_{\tilde \chi}$ becomes large
enough that finite momentum effects become
important [34,35].  (The symmetric or
anti-symmetric linear combinations of ${\tilde H_1}$ and
${\tilde H_2}$ are also favored configurations.)  For a pure
\bino the \neut-nucleon couplings are determined by
[4]
$$a_{p (n)} = \bigg\{{1\over{2 \cos^2 \theta_W}} \sum_{u,d,s} \Delta q [
(T_{3L} - e_q)^2 + e_q^2]\bigg\}\zeta_q\eqno(12)$$
$$ \zeta_q = -2 {m_W^2\over{m_{\tilde q}^2 - m_{\tilde \chi}^2}}
\sin^2\theta_W$$
where $\theta_W$ is the weak angle, $T_{3L}$ is the third
component of weak isospin, $e_q$ is the charge of the quark,
and $\Delta q$ is the fraction of the nucleon's spin carried by
each flavor of quark (see the appendix).
The sum is over the up, down, and strange
quarks in the nucleon. The term in the curly brackets
has often been referred to as $a_{p(n)}$ in much of the
previous literature [10,11,13].  Following ref. [36] the EMC
values for the proton $\Delta q$'s are: $\Delta u = 0.78, \; \Delta d 
= -0.50$, and $ \Delta s = -0.16$, all with an uncertainty of $\pm 0.08$.
Inserting these values into eq. (12) gives $a_p = 0.179\zeta_q$ and
$a_n = -0.097\zeta_q$, or
equivalently, $a_0 = 0.082\zeta_q$ and $a_1 = 0.276\zeta_q$.
The NQM values of the spin content are [37]
$\Delta u = 0.93, \; \Delta d = -0.33$, and $\Delta s = 0.0$
resulting in $a_p = 0.254\zeta_q$ and
$a_n = -0.017\zeta_q$ ($a_0 = 0.237\zeta_q$,
$a_1 = 0.271\zeta_q$).

In Fig. 5 we show $S(q)$ for \si29 using the
shell model calculations discussed in Section IIa.  The solid
line is computed using the EMC values of $a_{p(n)}$ and the dashed
line results from using the NQM values.  For \si29, the
response is dominated by the contribution from neutrons. Hence, the
small value of $a_n$ in the NQM leads to a severe depression
in the calculated matrix element.  In Fig. 6a we plot
$S(q)$ for \ge73 using the same set of $a$'s.  Here we see
similar behavior but with a much larger amplitude due to the
larger amount of spin carried by the neutrons in \ge73
$\{ (\langle {\bf S}_n\rangle_{Ge}/\langle {\bf S}_n\rangle_{Si})^2
= (0.47/0.13)^2 \simeq 13 \}$.  Clearly \ge73 is a better choice as
a detector for the pure \bino case.  Next, in Fig. 6b
we present $S(q)$ for \ge73 where we have quenched the isovector
coupling constant in accordance with the discussion of Section III,
$a_1^{\prime} = 0.833 a_1 = 0.230$. The quenching results in an
overall decrease in the amplitude of $S(q)$ for the EMC cross
section and a slight increase in the NQM counterpart.

Finally,
in Fig. 7 we present $S(q)$ for a photino scattering with
\ge73.  A photino massive enough to make finite momentum effects
relevant is highly unlikely [34,35].  However, Fig. 7 is
useful, both because a pure photino has often been considered
in the literature, and because it illustrates the effects upon
$S(q)$ of changing
$a_n$ and $a_p$. For a pure photino, the nucleon
couplings are
$$a_{p(n)} = \bigg\{\sum_{u,d,s} \Delta q e_q^2\bigg\}\zeta_q,
\eqno(13)$$
resulting in: $a_p = 0.273\zeta_q$ and $a_n = -0.153\zeta_q$ for the EMC
values of the spin content and $a_p = 0.376\zeta_q$
and $a_n = -0.043\zeta_q$ for
the NQM estimates.  Now
that we have presented our results, we may compare them to
other work.

In Fig. 8 we compare $S(q)$ for our \si29 result with the shell
model (solid line), the single particle model (dashed line),
and the single particle model quenched so that $S(0)$ agrees with
the OGM prediction (dash-dotted line).
This last is not self-consistent but it is the only
way to extend the OGM to finite momentum transfers [12].
Two things are immediately apparent
from Fig. 8.  First, the unquenched ISPSM grossly
over-predicts the value of $S(q)$ for all $q$.  Second, the ISPSM
with OGM quenching seems to do a very good job of
predicting $S(q)$.  We quench by reducing $a_n$ by the
factor $0.15/0.50 = 0.3$.  There is a slight difference in their
$q = 0$ normalization that is maintained throughout (see Table
1).  This mismatch would be exacerbated by using the value of
$\langle {\bf S}_n \rangle \simeq 0.2$ of the
EOGM [7,11].  There is also a very slight difference between
the shapes of the ISPSM and the shell model results but this appears
rather trivial.

Figure 9 is the \ge73 counterpart to Fig. 8.  Here,
$S(q)$ is plotted for:  the large space shell model results
with no quenching (solid line); the large space shell model
result with an isovector quenching factor of 0.833 (long
dashed line); the ISPSM with and without a similar isovector
quenching (dotted and short dashed lines respectively);
and the ISPSM quenched to agree with the OGM at $q = 0$
(dot-dashed line).  The most relevant comparison is between
the shell model results, quenched or unquenched, and the
ISPSM results quenched to agree with the OGM at $q = 0$
($a_n \rightarrow (0.23/0.5)a_n$).  We see that there is a very
large discrepancy of either a factor $\sim 2$ or $\sim 4$ for
the \bino case.  Configuration mixing has made
a major difference in this case because it has allowed for a large
contribution to $\langle {\bf L}_p \rangle$, see Table 1.
The quenched IBFM result, when combined with a ISPSM response
function, is in accord with the OGM + ISPSM result.

A final issue which bears mentioning, in both the zero and
finite momentum transfer regimes, is the uncertainties in the
nucleon coupling constants, $a_n$ and $a_p$.  So far, we have
concentrated upon our attempts to improve the estimates of the
cross section by generating more realistic nuclear wave
functions.  We have already demonstrated, in Figs. 5
and 6, the large differences
in the matrix element which result from the EMC vs. the NQM
estimates of the proton's spin content.
Here we wish to point out that even within the context of the
EMC result, large differences in the rate are allowed.
By varying the EMC values of the $\Delta q$'s within
the error bars we can vary $S(q)$ as shown in Fig. 10.  Here
the solid line is the same as the solid line in Fig. 9 and the
other lines are obtained by varying $\Delta u$, $\Delta d$, and
$\Delta s$ within the allowed EMC errors.  A huge range of values is
allowed for $S(q)$.  In fact, the uncertainty due to the
$\Delta q$'s is far larger than that arising due to nuclear
processes.  Hopefully, new experiments will eventually determine
the $\Delta q$'s with better precision.  These new values, when
coupled with the new picture of the nuclear structure presented
here, will allow for a more accurate and complete description of
\chiN scattering for any neutralino.

In this section we have demonstrated that
finite momentum transfers can have rather sizable effects
on the \chiN scattering
cross section and hence on the scattering rate.  In most cases,
heavier \neut's will have suppressed cross sections.  This,
coupled with the fact that it takes fewer heavy \neut's
than light ones to account for the dark matter,  implies
that it will be
more difficult to detect neutralinos via the
direct detection technique if they are very heavy.

\bigskip
\noindent{\bf V. Discussion and Conclusions}
\bigskip

In this paper we have presented detailed nuclear shell model
calculations of spin-dependent elastic \chiN scattering from
\si29 and \ge73.  Major \neut dark matter detection programs
based upon these two nuclei are well under way [15].   Our results
reveal both the strengths and the weaknesses of previous,
more phenomenological, analyses [7].  In both nuclei
the ISPSM is found to be a poor approximation to the true nuclear
configuration.  In \si29 at zero momentum transfer, the OGM finds
results which are in accord with those of our calculations.  This
shows that, for this nucleus, the OGM adequately represents the
configuration mixing. This is not the case
in \ge73.  In this nucleus, the shell model result is in severe
disagreement with the OGM and quenched IBFM results.  The main
difference comes from the large amount of orbital angular
momentum carried by the protons.  This large value for
$\langle {\bf L}_p \rangle$ violates a key assumption of the
OGM and is not found in the IBFM.  The larger value of
$\langle {\bf L}_p \rangle$ in our calculation leads to a
large improvement in the calculated vs. measured magnetic
moment of \ge73 but still does not lead to complete agreement with free
particle $g$-factors.  Thus
we have discussed the difficult issue of quenching as it
applies to \ge73.  Even with quenching included our results still
differ from those of the OGM and IBFM.

Unlike the OGM or the IBFM, it is straightforward to extend
the shell model results to finite momentum transfer in a
consistent manner.  We find that finite momentum transfer will
play an important role in determining the \chiN scattering rate for
very heavy neutralinos.  Here again, we have compared the
shell model results against the OGM combined with the ISPSM form
factor.  This latter approach seems to work well for \si29 but
significant disagreement is found for \ge73.

The major concern regarding our results is the validity of
the ground state wave functions we have generated.  To help
allay that concern, we have used our wave functions to calculate
a number of nuclear observables and compared the results
to the measured values.  In \si29 excellent agreement was found for
the magnetic moment, the excited state energy spectrum, and the
first 3 spectroscopic factors.  This was not surprising since
we were able to use a well tested interaction in a well
defined model space.  These calculations in \si29 serve another
useful purpose.  The disagreement between the calculations
and measurements in \si29 serve to calibrate how accurately
one can calculate nuclear properties in the shell model.  Thus,
the magnitude
of the ``errors'' in the good calculation of silicon will help
to establish how believable our calculation of \ge73 is.

In \ge73 the picture is more complicated. No phenomenological
interaction has been developed for this region of the
periodic table and fairly severe truncations
to the model space have to be
imposed to obtain manageable dimensions.
Despite these obstacles, our best calculated ground state wave
function for \ge73 seems to be a good match  when compared
to experiment.  We have obtained good agreement with the low
lying excited state energy spectrum.  The ground state to
ground state spectroscopic factor is also in fair
agreement with experiment.  The spectroscopic factors for
transitions to excited states reveal that our model space
does not contain enough configurations to completely describe
the excited states, although the agreement between theory and
experiment is not unreasonable.  We reiterate that our model
space was designed to describe the ground state and for that it
does an acceptable job.

The final comparison is between magnetic
moments.  Using the large space wave function a significant improvement
over the ISPSM or IBFM estimates of $\mu$ is found.  Unfortunately,
perfect agreement with experiment is not obtained but this
is not unexpected or particularly damaging.  Some sort of
quenching of both $\mu$ and the
GT spin matrix elements is almost universally required in shell model
calculations of heavy nuclei.  We find that an isovector
spin quenching factor of 0.833 results in agreement with the
measured value of the magnetic moment.  This method and
amount of quenching are consistent with previous studies of both
GT and $\mu$ quenching and is therefore well motivated and simple.
It is not manifestly obvious that quenching is needed in
\ge73--\neut scattering but, if so, we are confident that the
correct answer lies in the range between the quenched and
unquenched values of our calculation.
In light of the above comparisons we feel that we have
obtained an adequate description of the ground state structure
of \ge73 and that any calculation in a larger model space or with
a different interaction will confirm our results.
The disagreement between calculation and measurement in each
case is somewhat larger than for \si29 but is not so large
as to cause us to doubt the result.  The case is not airtight
but 3 different lines of argument all indicate that
the large space wave function is a better representation of the
real nucleus than has previously been available.

As a summary and ``User's Guide'' we have collected the formulae, references,
and suggested quenching methods in the appendix.  Formulae for the
spin-dependent neutralino cross section and the event detection
rate in a bolometric detector are also given there.

In Section IV we described the formalism for \chiN scattering
at finite momentum transfer.  The cross section is given by
eq. (7) and all of the nuclear physics is contained in the
spin structure function, $S(q)$, given by eqs. (8)--(10). $S(q)$
may be decomposed into a pure isovector piece $S_{11}$, a pure
isoscalar piece $S_{00}$, and an interference term $S_{01}$,
see eq. (10) (we will refer to these as $S_{ij}$ where
$i,j = 0,1$).  By taking suitable linear combinations of
$a_n$ and $a_p$, the isoscalar and isovector
nucleon-\neut coupling constants, $a_0$ and $a_1$,
can be found for any composition neutralino.  In Figs.
3 and 4 we presented the $S_{ij}$ for \si29 and \ge73
respectively.  Along with $a_0$ and $a_1$ these plots can
be used to compute the terms $a_i a_j S_{ij}$ which go into
eq. (10) to find $S(q)$ for any given \neut.  As an alternative
to this we have provided convenient fits to $S(q)$ for both nuclei
in eq. (11).  These fits are quite accurate, for most \neut
compositions, in the regime $y \la 0.15$.  As noted in section
IV, there are certain combinations of $a_0$ and $a_1$ for
which  eq. (11) is a poor approximation.
With these words as a guide and using the formulae in the appendix,
it should be straightforward to incorporate this
work into calculations of the \chiN scattering rate.

To help place this work in a larger context, we close by
briefly discussing one final issue.  This is a brief
comparison of the spin dependent vs. the spin {\bf in}dependent
scattering.  In addition to the spin dependent scattering
discussed at length in this paper, there is spin independent
scattering which may produce competitive or larger scattering
rates [4,7].  The spin independent piece is coherent,
and thus, the cross section scales as (atomic number)$^2$.  So,
for heavy nuclei or light Higgs bosons
this piece may dominate depending upon the mass and
composition of the neutralino.  Engel [7,18] has shown
that the spin-independent form factor, corresponding to $S(q)$,
falls off much faster at large $q$ than does $S(q)$ itself.  Thus,
for very large $m_{\tilde \chi}$ spin dependent scattering
is likely to dominate.  Furthermore, for the pure \neut states
(e.g. a pure \bino), the coherent piece is suppressed(note,
however, that recent work
by Nojiri and Drees [38] may change this conclusion).
Therefore spin-dependent scattering will probably dominate for
these pure states.  Outside of these two regimes it is not clear
which piece, spin dependent or independent, will dominate the
cross section and hence the rate.

\bigskip
\bigskip
\noindent{\bf Acknowledgements}
\bigskip

M.T.R. would like to thank his wife, Lisa, for assisting in a number
of the computations during her battle with cancer.  We would also
like to thank J. Engel for several useful discussions.  Work
at LLNL was performed under the auspices of the U.S. Department of Energy
under contract W-7405-ENG-48 and DOE Nuclear Theory Grant
SF-ENG-48.  K.G. acknowledges support from a DOE Outstanding Junior
Investigator grant.

\bigskip
\bigskip
\bigskip

\noindent{\bf Appendix: User's Guide}
\bigskip

In this appendix we collect the formulae necessary to convert
our results for the nuclear matrix elements into usable
cross sections and interaction rates for dark matter detectors.
We give {\it only} the spin-dependent cross section.  To get the
total cross section the spin-independent
cross section [4,5,39,38] must be added to the cross sections
below.
We follow the notation and conventions of references [4] and
[5] with only slight modifications.
In the limit of zero momentum transfer the cross section for spin-dependent
elastic \chiN scattering is
$$\sigma_{SD} = {{24 m_{\tilde \chi}^2 m_N^2 G_F^2}\over{\pi
(m_{\tilde \chi} + m_N)^2}} {4\over{3}} \Lambda^2 J(J + 1)
 \eqno(A1)$$
where
$m_N$ is the nuclear mass, $m_{\tilde\chi}$ is the neutralino mass,
$J$ is the total nuclear spin, $G_F$ is the Fermi constant, and
$$\Lambda = {{\langle N\vert a_p {\bf S}_p + a_n {\bf S}_n
\vert N \rangle}\over{J}} =
 {{a_p \langle {\bf S}_p \rangle}\over{J}}  +
{{a_n \langle {\bf S}_n \rangle}\over{J}}. \eqno(A2)$$
The spin averages
$\langle {\bf S}_p \rangle$ and $\langle {\bf S}_n \rangle$
are given in Table 3 (but see below  concerning
quenching before using them), while
$a_p$ and $a_n$ contain the details of the supersymmetric model, as
well as the quark spin content of the proton and neutron:
$$a_{p(n)} = \sum_{u,d,s} A^{\prime}_q \Delta q, \eqno(A3)$$
$$A^{\prime}_q = {1\over{2}} T_{3L}^q(Z_3^2 - Z_4^2) -
x_q^2 \bigg\{ {{2 m_q^2 d_q^2}\over{4 m_w^2}} + [T_{3L}^q Z_2 -
\tan \theta_W(T_{3L}^q - e_q) Z_1 ]^2 +
\tan^2\theta_W e_q^2 Z_1^2\bigg\}, \eqno(A4)$$
and
$$x_q^2 = {{m_W^2}\over{ m_{\tilde q}^2 -m_{\tilde \chi}^2 }}. \eqno(A5)$$
The various quantities used in the definition of
$A_q^\prime$ involve the neutralino
couplings and mass matrix, and can be found in ref [4].
For the special cases of photinos and B-inos, the $a_{p(n)}$
are given in eqs. (12) and (13).
The values $\Delta q$
of the spin content of the nucleons are not well known.  Some commonly
used values are given in the text just after eq. (12).
The $\Delta q$ for the neutron can be found from the $\Delta q$ for
the proton by switching $\Delta u$ and $\Delta d$.
(We note, for completeness, that new data on the deuteron's
spin structure function has very recently become available.  A
new analysis incorporating this data along with the EMC results
yields values of
$\Delta u = 0.81$, $\Delta d = -0.45$, and $\Delta s = -0.12$;
all with an uncertainty of $\pm 0.03$ [41].)

For \si29, we suggest using the values of $\langle {\bf S}_n \rangle$
and $\langle {\bf S}_p \rangle$ from the ``Shell Model" row of
Table 3 with no quenching.
For \ge73, we advocate quenching the
isovector piece of the nucleon-\neut coupling constants
$a_p$ and $a_n$.  Hence one should not use
eq. (2).
To quench the isovector piece of $a_p$ and $a_n$ one first
computes $a_p$ and $a_n$ from eq. (A3) and then constructs
the isovector and isoscalar coupling constants:
$a_1 = a_p - a_n$ and $a_0 = a_p + a_n$. The isovector
piece is then multiplied by the quenching factor,
$a_1^{\prime} = 0.833 a_1$.  In the zero momentum transfer
limit, the $a_p$ and $a_n$ are then recalculated using the
quenched value $a_1^{\prime}$.
These values are then used in
eq. (A2) in place of those calculated directly from eq. (A3).
In other words to quench by a factor $Q$ ($ = 0.833$) in eq. (A2) make the
replacement:
$$
a_p \Longrightarrow a_p^\prime =
{1\over 2}[ a_p(1+Q) + a_n(1-Q)] = 0.917a_p + 0.0835a_n
$$
$$
a_n \Longrightarrow a_n^\prime =
{1\over 2}[ a_p(1-Q) + a_n(1+Q)] =  0.0835a_p + 0.917a_n.
\eqno(A6)
$$
In this scheme one uses
the values of $\langle {\bf S}_n \rangle$
and $\langle {\bf S}_p \rangle$ from the ``Shell Model (large)" row of
Table 3.  (not the ``quenched" row.)

To make contact with previous work, we note that the ISPSM
(single particle shell model) can be obtained from eq.~(A1)
by the simple substitution
$$ \Lambda^2 \Longrightarrow a_{p(n)}^2 \lambda^2,
\eqno(A7)$$
where in the ISPSM only one of either $a_n$ or $a_p$
contributes and
$$\lambda = {1\over{2}} \bigg[1 + {{s_{p(n)}(s_{p(n)} + 1) -
l_{p(n)} (l_{p(n)} + 1)}\over{J(J + 1)}}\bigg] = {s_{p(n)}\over{
J}} \eqno(A8)$$
where $J$ is the total nuclear angular momentum $(J =
s_{p(n)} + l_{p(n)})$, $s_{p(n)}$ and
$l_{p(n)}$ are the unpaired proton or neutron spin and
orbital angular momentum respectively.
For example, in the ISPSM, \ge73 is modeled as a neutron
with $l_n = 4$ and $J={9\over2}$,
while \si29 is modeled as a neutron with
$l_n = 0$, and $J={1\over2}$.

When the neutralino mass is over a few tens of GeV, it becomes important
to include the effect of finite momentum transfer.
Eqs.~(7-11) in the text show how to include this effect.
To implement the suggested quenching scheme for \ge73 simply
make the replacement eq.~(A6) (or equivalently $a_1 \rightarrow
a_1^\prime = 0.833 a_1$) in eqs.~(7-11).
Note in particular that eq.~(9c) can be written:
$$S(0) = {2 J + 1\over{\pi}} \Lambda^2 J(J + 1). \eqno(A9)$$
Since we work in the $z$-projection of
the maximally stretched state, the function $f(N)$ introduced in
eq. (9c) has the form $f(N) = f(J) = (2 J + 1)(J + 1)/J$.

The rate per unit detector mass
with which neutralinos interact with a detector is given by
[9,4,7]
$$R = {\rho\over{m_{\tilde \chi}m_N}}
\int v dv f(v) \int_0^{4M_r^2v^2} dq^2
{d\sigma\over{dq^2}}, \eqno(A10)$$
where,
$m_N$ is the mass of the nucleus,
$M_r = m_N m_X/(m_N+m_X)$, $d\sigma/dq^2$ is given in eq.~(7),
$\rho \simeq 0.4 \GeV/\cm^3$ is the halo density (assumed to
consist entirely of neutralino's), $f(v)$ is the velocity distribution
of the halo, and
$v_0 \simeq 220$km s$^{-1}$ is
the circular velocity in the halo, which sets
the scale for the halo gravitational potential.
For simplicity one  can choose a
a Maxwellian distribution of velocities
$f(v) = 4v^2\exp{(-v^2/v_0^2)}/(v_0^3\sqrt{\pi})$, or one can follow
[9,4,40] and include the effects of the Sun and Earth motion, as well
as the effect of a detector threshold.

Integrating eq.~(A10) one gets
$$R = {2\over{\sqrt\pi}} \eta_c\eta_v
 {\rho\over{m_{\tilde \chi}m_N}} v_0
\sigma_{SD}, \eqno(A11)$$
where [9] $\eta_v \approx 1.3$ gives a correction for the speed of
the Sun through the halo, and the effect of integrating over
$q^2$ is summarized by
$$
\eta_c = \sigma_{SD}^{-1}\int dq^2 {d\sigma\over dq^2}.
$$
In the limit of zero momentum
transfer, the $q^2$ integral in
eq. (A1) becomes $\sigma_{SD}$ and $\eta_c \rightarrow 1$.
Note again that to find the total event rate one must include the
spin-independent piece of the cross section which is not considered
in this paper.

In order to evaluate $\eta_c$ one may use the
fits to $S(q)$ given in eq. (11).
This will give an approximation which should be good
for most neutralinos (but see the cautions in the text).
We re-write eq. (11) as
$$S^{fit}(y) = \sum_{i=1}^3  A_i e^{-B_iy},$$
where the $A_i$ and $B_i$ are given in eq.~(11), and $y=(bq/2)^2$.
For \ge73, $b \simeq 2.04$fm, and for \si29, $b\simeq1.75$ fm.
Note now $\sum A_i = S^{fit}(0)$.
Then
$$
\eta_c = \left\{ \sum_i A_i { 1-\beta_i \over \alpha_i (
{1\over2} + p^2 + {pe^{-p^2} \over \sqrt{\pi} \erf p})} \right\}
/ \sum_i A_i,
\eqno(A12)
$$
where
$$
\beta_i = {\exp{-p^2\alpha_i \over 1+\alpha_i} \erf{p \over
\sqrt{1+\alpha_i}} \over \sqrt{1+\alpha_i} \erf p},
$$
$p = v_{SUN}/v_0 \approx 1$, and $\alpha_i =
B_ib^2 v_0^2M_r^2$.
In the limit $p=0$ (the limit
of a stationary Sun or isotropic velocity distribution)
$$
\eta_c \rightarrow \sum_i A_i \alpha^{-1}[1 -1/(1+\alpha_i)] /
\sum_i A_i.
$$

\vfill\eject

\noindent{\bf References}
\medskip

\item{1.} V. Trimble, {\it Ann. Rev. Astron. Astrophys.} {\bf 25},
425 (1989); J.R. Primack, B. Sadoulet, and D. Seckel, {\it Ann. Rev.
Nucl. Part. Sci.} {\bf 38}, 751 (1988).

\item{2.} M.S. Turner, {\it Phys. Scripta} {\bf T36}, 167 (1991);
J. Ellis, {\it Phys. Scripta} {\bf T36}, 142 (1991).

\item{3.} H.E. Haber and G.L. Kane, {\it Phys. Rep.} {\bf 117}, 75 (1985).

\item{4.} K. Griest, {\it Phys. Rev.} {\bf D 38}, 2357 (1988).

\item{5.} J. Silk, K.A. Olive, and M. Srednicki, {\it Phys. Rev. Lett.}
{\bf 53}, 624 (1985);  M. Kamionkowski, {\it Phys. Rev.} {\bf D44}, 3021
(1991).

\item{6.} M.W. Goodman and E. Witten, {\it Phys. Rev.} {\bf D31},
3059 (1985).

\item{7.} J. Engel, S. Pittel, and P. Vogel, {\it Int. J. Mod. Phys.}
{\bf E 1}, 1 (1992).

\item{8.} J. Ashman, {\it et al.}, {\it Phys. Lett.} {\bf B206}, 364 (1988);
J. Ashman, {\it et al.}, {\it Nucl. Phys. } {\bf B328}, 1 (1989).

\item{9.} A.K. Drukier, K. Freese, and D.N. Spergel, {\it Phys. Rev.}
{\bf D 33}, 3495 (1986); K. Freese, J. Freeman, and A. Gould, {it Phys.
Rev.} {\bf D 37}, 3388 (1988).

\item{10.} J. Ellis and R.A. Flores, {\it Nuc. Phys.} {\bf B 307},
883 (1988).

\item{11.} J. Engel and P. Vogel, {\it Phys. Rev.} {\bf D40}, 3132
(1989).

\item{12.} J. Ellis and R.A. Flores,  {\it Phys. Lett.} {\bf B263},
259 (1991).

\item{13.} A.F. Pacheco and D. Strottman, {\it Phys. Rev.} {\bf D40},
2131 (1989).

\item{14.} J. Engel, {\it et al.}, {\it Phys. Lett.} {\bf B275}, 119 (1992).

\item{15.} B.A. Young {\it et al.}, {\it Nucl. Inst. Meth.}
{\bf A311}, 195 (1992); T. Shutt, {\it et al.}, {\it Phys. Rev. Lett.}
{\bf 69}, 3425 (1992); {\bf 69}, 3531 (1992); B.A. Young in {\it Proceedings
of the Third International Conference on Advanced Technology
and Particle Physics}, ed. E. Borchi {\it et al.}(Elsevier, Amsterdam, 1993).

\item{16.} D.A. Resler and S.M. Grimes, {\it Computers in Phys.}
{\bf 2} (3), 65 (1988).

\item{17.} F. Iachello, L.M. Krauss, and G. Maino, {\it Phys. Lett.}
{\bf B254}, 220 (1991).

\item{18.} J. Engel {\it Phys. Lett.} {\bf B264}, 114 (1991).

\item{19.} J.D. Walecka, in {\it Muon Physics}, eds. V.W. Hughes and
C.S. Wu, Vol. 2 (Academic Press, New York, 1975) p. 113;  T.W. Donnelly
and R.D. Peccei {\it Phys. Rept.} {\bf 50}, 1 (1979; T.W. Donnelly and
W.C. Haxton, {\it Am. Data Nucl. Data Tables.} {\bf 23}, 103 (1979).

\item{20.} B.H. Wildenthal, {\it Progress in Particle and Nuclear
Physics} {\bf 11} ed. D.H. Wilkinson (Pergamon  Press, Oxford, 1984),
p.5.

\item{21.} B.A. Brown, {\it Nucl. Phys.} {\it A522}, 221c (1991).

\item{22.} C.M. Lederer and V.S. Shirley, {\it Table of Isotopes,
7th ed.} (John-Wiley, New York, 1978).

\item{23.} K.L.G. Heyde, {\it The Nuclear Shell Model} (Springer-Verlag,
Berlin, 1990) p.226; P.J. Brussaard and P.W.M. Glaudemans,
{\it Shell-Model Applications in Nuclear Spectroscopy}
(North-Holland, Amsterdam, 1977) p.144.

\item{24.} P.M. Endt {\it Nucl. Phys.} {\bf A521}, 1 (1990).

\item{25.} G.J. Mathews, {\it et al.}, {\it Phys. Rev.} {\bf C 32},
796 (1985).

\item{26.}  F. Petrovich, {\it et al.}, {\it Phys. Rev. Lett.} {\bf 22},
895 (1969).

\item{27.} A. Kallio and K. Kolltveit, {\it Nucl. Phys.} {\bf 53},
87 (1964).

\item{28.} B.A. Brown and B.H. Wildenthal, {\it Nucl. Phys.}
{\bf A474}, 290 (1987).

\item{29.} W.A. Richter, {\it et al.}, {\it Nucl. Phys.}
{\bf A523}, 325 (1991).

\item{30.} W.A. Yoh, S.E. Darden, and S. Sen, {\it Nucl.
Phys.} {\bf A263}, 419 (1976).

\item{31.} B.A. Brown and B.H. Wildenthal, {\it At.
Data Nucl. Data Tables.} {\bf 33}, 347 (1985).

\item{32.} J. Ellis, {\it et al.}, {\it Nucl. Phys.} {\bf B238}, 453 (1984).

\item{33.} K. Griest and L. Roszkowski, {\it Phys. Rev.} {\bf D 46},
3309 (1992).

\item{34.} K Griest, M. Kamionkowski, and M.S. Turner, {\it Phys. Rev.}
{\bf D 41}, 3565 (1990);  K.A. Olive and M. Srednicki {\it Nucl.
Phys.} {\bf B355} 208 (1991).

\item{35.} L. Roszkowski, Univ. of Michigan Preprint UM-TH-9306.

\item{36.} R.L. Jaffe and A. Manohar, {\it Nucl. Phys.} {\bf B337},
509 (1990).

\item{37.} R. Flores, K.A. Olive, and M. Srednicki, {\it Phys.
Lett.} {\bf B 237}, 72 (1990).

\item{38.} M. Drees and M.M. Nojiri, {\it Phys. Rev.}
{\bf D}, 4226 (1993).

\item{39.} R.Barbieri, M.Frigeni, and G.G.Guidice, {\it Nucl. Phys.}
{\bf B313}, 725 (1989).

\item{40.} K.~Griest, {\it Phys. Rev.} {\bf D 37}, 2703 (1988).

\item{41.} J. Ellis and M. Karliner, CERN Preprint, CERN-TH-6898/93
(1993).

\vfill\eject

\centerline{{\bf Table 1.}
Summary of the calculated spin of $^{35}$Cl.}
\smallskip
$$\vbox{
\hrule\smallskip
\hrule\smallskip
\halign { \hfil #\hfil \ \ & \ \ \hfil  #\hfil \ \ &
 \ \ \hfil  #\hfil \ \  & \
\hfil #\hfil \ \  \cr
 & $\langle{\bf S}_n \rangle$ & $\langle{\bf S}_p \rangle$ \cr
 \noalign{\smallskip}\noalign{\hrule}\noalign{\smallskip}
 ISPSM & 0 & -0.3 \cr
 OGM [7,11] & 0 & -0.15 \cr
 EOGM [7,11] & 0.014 & -0.094 \cr
 Shell Model [13] & -0.011 & -0.059 \cr
 This Work & -0.0088 & -0.051 \cr
 \cr
}\hrule}$$

\bigskip
\bigskip
\bigskip

\noindent{{\bf Table 2.}
Summary of the spectroscopic factors, $S(J;jJ_0)$ for $^{29}$Si.  The
experimental values are an unweighted average of the data followed
by (in parentheses) the spread in the data, as represented by
the low and high measurements, from ref. [24].}
\smallskip
$$\vbox{
\hrule\smallskip
\hrule\smallskip
\halign { \hfil #\hfil \ \ & \ \ \hfil  #\hfil \ \ &
 \ \ \hfil  #\hfil \ \  & \
\hfil #\hfil \ \ & \ \ \hfil  #\hfil \ \  \cr
 Orbital & $J^{\pi}$ & $(2J + 1) S_{Exp.}$ & $(2J + 1) S_
{Calc.}$ & $(2J + 1) S_{ISPSM}$ \cr
 \noalign{\smallskip}\noalign{\hrule}\noalign{\smallskip}
 $2s_{1/2}$ & ${1\over{2}}^+$ & 1.02 (0.74--1.5) & 0.90 & 2 \cr
 $1d_{3/2}$ & ${3\over{2}}^+$ & 3.62 (2.1--5.2) & 2.65 & 4 \cr
 $1d_{5/2}$ & ${5\over{2}}^+$ & 1.14 (0.53--1.8) & 0.85 & 6 \cr
 \cr
}\hrule}$$

\bigskip
\vfill\eject

\noindent{{\bf Table 3.}  The value of the (z-projection) of the
nuclear spin and orbital angular momentum matrix elements and the
predicted magnetic moments, $\mu$, (using free particle
$g$-factors) for $^{29}$Si and $^{73}$Ge. The last row contains
our values of the large space shell model calculation with an
isovector quenching factor of 0.833 applied to the spin matrix
elements of the row above.  If the
magnetic moment has been used as input it is necessarily the measured
value and has been enclosed in parentheses.
}
\smallskip
$$\vbox{
\hrule\smallskip
\hrule\smallskip
\halign { \hfil #\hfil \ \ & \ \ \hfil  #\hfil \ \ &
\ \ \hfil  #\hfil \ \ &
\ \ \hfil  #\hfil \ \ &
 \ \ \hfil  #\hfil \ \  & \
\hfil #\hfil \ \  \cr
 & $\langle {\bf S}_n \rangle$ & $\langle {\bf S}_p \rangle$ &
$\langle {\bf L}_n \rangle$ & $\langle {\bf L}_n \rangle$ & $\mu$
\cr
 \noalign{\smallskip}\noalign{\hrule}\noalign{\smallskip}
 $^{29}$Si & & & & & \cr
 \noalign{\smallskip}\noalign{\hrule}\noalign{\smallskip}
 ISPSM & 0.5 & 0 & 0 & 0 & -1.91 \cr
 OGM & 0.15 & 0 & 0.35 & 0 & (-0.555) \cr
 EOGM & 0.204 & 0.054 & ? & ? &  (-0.555)\cr
 Shell Model & 0.13 & -0.002 & 0.37 & 0.019 & -0.50 \cr
 \noalign{\smallskip}\noalign{\hrule}\noalign{\smallskip}
 $^{73}$Ge & & & & & \cr
 \noalign{\smallskip}\noalign{\hrule}\noalign{\smallskip}
 ISPSM & 0.5  & 0  & 4 & 0 & -1.91 \cr
 OGM & 0.23 & 0 & 4.27 & 0 & (-0.879) \cr
 IBFM & 0.469 & -0.009 & 3.981 & 0.060 & -1.785 \cr
 IBFM(quenched) & 0.245 & -0.005 & ? & ? & (-0.879)  \cr
 Shell Model (small) & 0.496 & 0.005 & 3.596 & 0.40 & -1.468 \cr
 Shell Model (large) & 0.468  & 0.011 & 3.529 & 0.491 & -1.239 \cr
 Shell Model (large,quenched) & 0.372  & 0.009 & 3.529 & 0.491 & (-0.879)
 \cr
}\hrule}$$

\bigskip

\noindent{{\bf Table 4.}
Summary of the spectroscopic factors, $S(J;jJ_0)$ for $^{73}$Ge.  The
experimental values are an unweighted average of the data followed
by (in parentheses) the spread in the data, as represented by
the low and high measurements, from ref. [30].}
\smallskip
$$\vbox{
\hrule\smallskip
\hrule\smallskip
\halign { \hfil #\hfil \ \ & \ \ \hfil  #\hfil \ \ &
 \ \ \hfil  #\hfil \ \  & \
\hfil #\hfil \ \ & \ \ \hfil  #\hfil \ \  \cr
 Orbital & $J^{\pi}$ & $(2J + 1) S_{Exp.}$ & $(2J + 1) S_
{Calc.}$ & $(2J + 1) S_{ISPSM}$ \cr
 \noalign{\smallskip}\noalign{\hrule}\noalign{\smallskip}
 Large Space & & & \cr
 \noalign{\smallskip}\noalign{\hrule}\noalign{\smallskip}
 $1g_{9/2}$ & ${9\over{2}}^+$ & 6.16 (5.04--7.35) & 3.96 & 10 \cr
 $2d_{5/2}$ & ${5\over{2}}^+$ & 0.29 (0.28--0.30) & 0.74 & 6 \cr
 $2p_{1/2}$ & ${1\over{2}}^-$ & 0.67 (0.52--0.78) & $3.3\times 10^{-4}$ & 2 \cr
 \noalign{\smallskip}\noalign{\hrule}\noalign{\smallskip}
 Small Space & & & \cr
 \noalign{\smallskip}\noalign{\hrule}\noalign{\smallskip}
 $1g_{9/2}$ & ${9\over{2}}^+$ & 6.16 (5.04--7.35) & 5.91 & 10 \cr
 $2d_{5/2}$ & ${5\over{2}}^+$ & 0.29 (0.28--0.30) & 3.55 & 6 \cr
 $2p_{1/2}$ & ${1\over{2}}^-$ & 0.67 (0.52--0.78) & $\sim 0$  & 2
 \cr
}\hrule}$$

\bigskip
\vfill\eject

\noindent{{\bf Table 5.}  The sets of quenched $g$-factors used
in ref. [29] and their effect upon the magnetic moment of $^{73}$Ge
using the large space wave functions.
}
\smallskip
$$\vbox{
\hrule\smallskip
\hrule\smallskip
\halign { \hfil #\hfil \ \ & \ \ \hfil  #\hfil \ \ &
\ \ \hfil  #\hfil \ \ &
\ \ \hfil  #\hfil \ \ &
 \ \ \hfil  #\hfil \ \  & \
\hfil #\hfil \ \  \cr
 Interaction & $g_n^s$ & $g_p^s$ & $g_n^l$
&  $g_p^l$ & $\mu$
\cr
 \noalign{\smallskip}\noalign{\hrule}\noalign{\smallskip}
 Free & -3.826 & 5.586 & 0 & 1 & -1.239 \cr
 FPD6 & -3.41 & 6.75 & 0.05 & 0.71 & -0.995 \cr
 FPMI3 & -2.53 & 5.87 & -0.11 & 0.87 & -1.079 \cr
 KB1 & -3.19 & 4.87 & 0 & 1 & -0.948 \cr
 Experiment & & & & & -0.879
\cr
}\hrule}$$

\vfill\eject

\noindent{\bf Figure Captions}

\item{\bf Figure 1}  The calculated (left) and measured [22]
(right) excited state energy spectrum of \si29.  The ground state has
$J^{\pi} = {1\over{2}}^+$.  The dashed lines connect states with the
same $J^{\pi}$. The $J^{\pi}$ values for most of the states have been
included for reference.

\item{\bf Figure 2} The calculated small space (left), measured
[22] (center), and calculated large space (right) energy
spectra of \ge73.   The ground state has $J^{\pi} = {9\over{2}}^+$.
Where we have felt comfortable with the identifications, we have
connected states with identical  $J^{\pi}$'s by dashed lines.
The $J^{\pi}$'s for some of the lower lying states have been
noted on the plot.

\item{\bf Figure 3} The calculated functions $S_{00},\; S_{11}$, and
$S_{01}$ from eq. (10) for \si29 as a function of $y$.
The solid line =  $S_{00}$, the
short dashed line = $S_{11}$, and the long dashed line = $S_{01}$.

\item{\bf Figure 4} The calculated functions $S_{00},\; S_{11}$, and
$S_{01}$ from eq. (10) for \ge73 (large space) as a function of $y$.
The solid line =  $S_{00}$, the
short dashed line = $S_{11}$, and the long dashed line = $S_{01}$.

\item{\bf Figure 5} The spin structure function, $S(q)$, for a
pure \bino scattering with \si29.  The solid line is derived
using the EMC values for $a_{n(p)}$; the dashed line uses the
NQM estimates of these couplings.

\item{\bf Figure 6} A) The spin structure function, $S(q)$, for a
pure \bino scattering with \ge73 (large space).  The solid line is derived
using the EMC values for $a_{n(p)}$; the dashed line uses the
NQM estimates for these couplings.  B) The same as (A) but with
an isovector quenching factor of 0.833 applied to enforce
agreement with the measured magnetic moment at $q = 0$
($a_1 \rightarrow 0.833 a_1$).

\item{\bf Figure 7} The spin structure function, $S(q)$, for a pure
${\tilde \gamma}$ scattering with \ge73 (large space calculation).
The solid line uses the EMC values for $a_{n(p)}$, the long dashed
line uses these values but with the isovector piece quenched
by a factor of 0.833, and the short dashed line uses the NQM
estimates for $a_{n(p)}$.

\item{\bf Figure 8} The spin structure function, $S(q)$, for a
pure \bino (EMC couplings) scattering with \si29 for three
different models.  The solid line is the shell model result
and corresponds to the solid line in Fig. 5.  The dashed line is
the ISPSM result.  The dot-dashed line is the ISPSM structure
function with $a_n$ quenched so that $S(0)$ agrees with the
OGM result.

\item{\bf Figure 9} Several different models of $S(q)$ for a
pure \bino scattering with \ge73:  solid line = large space shell
model result with no quenching (= solid line in Fig. 6A),
long-dashed line = large space shell model result with isovector
quenching (= solid line in Fig. 6B), short-dashed line = ISPSM
result, dotted line = ISPSM result with isovector quenching, and
the dot-dashed line = ISPSM result quenched to agree with the OGM
at $q = 0$.

\item{\bf Figure 10} $S(q)$ for a pure \bino, with EMC couplings,
scattering with \ge73 (large space calculation).  Here we have
varied the $\Delta u$,  $\Delta d$, and  $\Delta s$ of the EMC
result within the allowed errors ($\pm 0.08$) in deriving
$a_{n(p)}$.  The uncertainty due to the $\Delta q$'s is currently
much larger than the nuclear uncertainty; compare with
figure 9.

\vfill\eject\end